# Behavioral Analytics for Continuous Insider Threat Detection in Zero-Trust Architectures

**Gaurav Sarraf**
Independent researcher

*Abstract*—Insider threats are a particularly tricky cybersecurity issue, especially in zero-trust architectures (ZTA) where implicit trust is removed. Although the rule of thumb is never trust, always verify, attackers can still use legitimate credentials and impersonate the standard user activity. In response, behavioral analytics with machine learning (ML) can help monitor the user activity continuously and identify the presence of anomalies. This introductory framework makes use of the CERT Insider Threat Dataset for data cleaning, normalization, and class balance using the Synthetic Minority Oversampling Technique (SMOTE). It also employs Principal Component Analysis (PCA) for dimensionality reduction. Several benchmark models, including Support Vector Machine (SVM), Artificial Neural Network (ANN), and Bayesian Network (Bayes Net), were used to develop and evaluate the AdaBoost classifier. Compared to SVM (90.1%), ANN (94.7%), and Bayes Net (94.9), AdaBoost achieved higher performance with a 98.0% ACC, 98.3% PRE, 98.0% REC, and F1-score (F1). The Receiver Operating Characteristic (ROC) study, which provided further confirmation of its strength, yielded an Area Under the Curve (AUC) of 0.98. These results prove the effectiveness and dependability of AdaBoost-based behavioral analytics as a solution to reinforcing continuous insider threat detection in zero-trust settings

## I. INTRODUCTION

The term "cybersecurity" describes the measures used to safeguard private and sensitive information stored in computer networks, as well as the systems that connect to them. It is generally accepted that real time control of cyber risks is more effective than eventual control. Insider threats have turned out as one of the hardest to detect currently and they are launched by individuals who have rightful access to systems [1]. Insider threats classified into three categories: malicious users who deliberate damage, oblivious users who unknowingly harm security through their lack of attention to policies, and external actors who use their authorized credentials to get access to stored sensitive information [2]. Trusted insiders can share confidential information, compromise security mechanisms, steal and sabotage resulting in reputational, financial and operational losses.

The insider threats may be present even in the context of Zero Trust Security (ZTS), as attackers can use lawful credentials, simulate the normal user behaviour, or circumvent weak controls. Identity-based and situation-based access control are the main tenets of Zero Trust Security (ZTS). Implementation of ZTS at scale, however, has its own challenges [3], among which are striking a balance between high security and usability, legacy systems integration, access data processing, and also to maintain efficiency in the system without creating bottlenecks in the systems. These issues emphasize the fact that ZTS decreases risk significantly, but insider threats are a major issue that needs to be constantly observed and properly detected [4][5].

Identity and Access management (IAM) aids ZTS by identifying the identity and role-based access, but traditional IAM systems are not effective enough in identifying subtle behavioural anomalies [6]. In this case, two aspects are essential in boosting security, which are AI and ML. AI and ML can detect malicious or dangerous insider behaviours by examining behavioural data (i.e., time of logins, device use, network interactions, access logs, etc.) to detect anomalies [7]. Machine learning models are able to keep up with changing user behaviours and enhance detection rates and false positives. Machine learning models can equally match the evolving user behaviour and optimize the detection rates and false positives. ML plays a vital role in the implementation of security [8]. AI and ML are able to identify irregularities in behavioural patterns, such as when a person logs in, device usage, network usage, and access pattern, that may indicate malicious or high-risk insider activity [9]. However, machine learning also possesses some advantages: it can automatically identify complex patterns, can evolve as behavior evolves, false alarms are reduced, and can be easily applied to large data sets. ML is highly beneficial for a constantly evolving Zero Trust environment's insider threat identification because of all these features [10].

The inclusion of AI and ML in ZTS make IAM a dynamic risk-sensitive system, rather than a fixed authentication tool. Behavioural analytics enable organizations to keep track of user activity and identify abnormal behaviours and act ahead of insider threats [11]. This strategy enhances security in general, identity should act as the main perimeter, and it is consistent with the principle of Zero Trust, according to which no user or device can be implicitly trusted even in a trusted environment.

### A. Motivation and Contributions of the Study

This study is driven by the increasing number of insider threats in organizations, when trusted persons abuse their access to undermine security. In the context of zero-trust architectures, where no entity is inherently trusted, continuous monitoring and detection of insider threats become critical to safeguarding sensitive information and systems. The traditional detection solutions tend to have problems with an uneven data set, high-dimensional behavioural information, and the insidiousness of insider behaviours. This research addresses these challenges by employing behavioural analytics integrated with machine learning to enhance detection accuracy and reliability. By uncovering hidden patterns in user activities, the proposed framework supports proactive defence mechanisms and strengthens the resilience of zero-trust environments. The main findings from this research are these:

- The CERT Insider Threat Dataset, a well-known standard for research on identifying insider threats, was used.
- Performed data cleaning to remove noise, inconsistencies and outliers.
- Used SMOTE, or the Synthetic Minority Oversampling Technique, to rectify class disparities and increase minority participation.
- Implemented normalization to standardize feature values and ensure uniformity across variables.
- Employed Principal Component Analysis (PCA) for dimensionality reduction, capturing essential behavioural features while minimizing redundancy.
- Developed an AdaBoost-based classification model to improve detection of malicious insider behaviour.

- Verified the model's efficacy and reliability through testing it with F1, REC, PRE, and ACC and PRE as metrics.

### B. Justification and Novelty of the Paper

The justification for this study lies in the increasing sophistication of insider threats that exploit trust boundaries, even within zero-trust security models. Traditional IAM systems fail to detect subtle behavioural deviations, necessitating advanced solutions. The novelty of this work stems from integrating behavioural analytics with robust pre-processing—SMOTE for class imbalance, PCA for dimensionality reduction, and AdaBoost for adaptive classification—applied to the CERT Insider Threat dataset. Unlike conventional approaches, this framework not only enhances detection accuracy but also ensures scalability and adaptability, offering a practical, AI-driven solution for continuous insider threat monitoring in zero-trust architectures.

### C. Organization of the Paper

The following is the paper's structure The section II discusses previous research on detecting insider threats. Data preparation, feature engineering, and model building are all detailed in Section III of the suggested technique. In Section IV, show the AdaBoost classifier's experimental findings and performance review. Section V discusses the limitations and outlines future directions for enhancing insider threat detection. Section VI concludes the study by summarizing key findings and contributions.

## II. LITERATURE REVIEW

Recent researches point out that insider threats use privileged access, which is challenging to detect. The deep learning method, behavioural analytics, and ensemble machine learning techniques have proven useful in modelling the user behaviour and, as a result, allow early identification of malicious insiders in the zero-trust architectures.

Tamanna (2020) explores the concept of insider threats, which deal with security dangers that originate from trusted individuals within an organization. Because insiders know sensitive information and system weaknesses, these threats are hard to detect. An approach to detect unusual user behaviour is introduced in the study. It makes use of DNN techniques, particularly a blend of CNN and LSTM. Implemented on a publicly available dataset (CMU CERT version r4.2), the model outperforms conventional ML methods in detecting insider risks, with a ROC of 0.914.[12]

Le, Zincir-Heywood and Heywood (2020) investigate the difficulties in identifying malevolent insider threats, such as imbalanced data and changes in behaviour. For user-centred insider threat identification, they suggest a machine learning system that can spot both insiders and harmful behaviours. Even when trained with sparse data, their tests show that the system can identify malevolent insiders at an 85% rate with a tiny FPR of 0.78%. Notably, it can detect malicious actions within 14 minutes. The comprehensive reporting enhances analytical insights into insider threat investigations.[13]

Singh, Mehtre and Sangeetha (2019) paper addresses the rising threat of insider attacks within organizations. Analysing human behaviour to detect harmful behaviours is emphasised, as is the critique of current security solutions that mostly concentrate on external threats. Using an ensemble hybrid algorithm that combines CNNs with MSLSTM, the authors introduce a method for user behaviour profiling that can detect patterns of anomaly in time series data. The findings show that MSLSTM is more effective than the conventional models, giving the AUC of 0.9042 and 0.9047 on training and test respectively, which is the successful operation to identify insider threats by using publicly available datasets [14]

Hu et al. (2019) paper talk about how easy insider attackers can use intranet systems because of unauthorized access. A novel user authentication system is proposed, which employs deep learning and mouse biobehavioural features to provide continuous identity verification and thwart insider attacks. By conducting experiments with an open-source dataset consisting of 10 users, it was shown that this method is capable of authenticating people in approximately 7 seconds. The FNR was 2.28 and the false acceptance rate was 2.94, indicating that it is efficient [15]

Hall et al. (2018) paper present the current issue of insider threats in information protection, which suggests an unresolved gap between the needs and solutions in the community. Utilizing the CERT dataset r4.2, the study employs machine learning classifiers to predict malicious insider scenarios, specifically the act of uploading sensitive information to WikiLeaks prior to leaving an organization. The work introduces a meta-classifier that enhances predictive performance beyond individual models and lays out a methodology for first-pass user summaries from organizational log data. To make it more precise, methods like boosting are used. The meta-classifier attained an ACC of 96.2% according to the confusion matrix and area under the ROC curve [16]

Gamachchi and Boztas (2017) explain the grave dangers posed by dishonest individuals who could appear as reliable sources while wreaking havoc on the nation's finances, security, and public and private organisations' reputations. To combat insider threats, it is essential to investigate user, system, and network aspects in depth because malicious operations are unexpected and ever-changing. This paper proposes an insider threat detection system that uses attributed graph clustering methods and an outlier ranking system to identify suspicious users within an organisation. With a receiver operating characteristic curve value of 0.7648, the empirical results demonstrate that the framework is successful and has great detection capabilities [17]

Table I presents a summary of recent works on hybrid deep learning, ensemble ML, and behavioural analytics on insider hreat data, which shows high detection ACC and overcomes such challenges as scalability, real-time implementation

TABLE I.　　RECENT STUDIES ON INSIDER THREAT DETECTION TECHNIQUES

| Reference | Methodology | Dataset | Results | Challenges | Future Work |
|---|---|---|---|---|---|
| Tamanna (2020) | Hybrid deep learning using LSTM + CNN for anomalous user behavior detection | CMU CERT r4.2 (12GB) | ROC = 0.914; outperforms traditional ML approaches | High computational cost for large-scale data; difficulty in real-time deployment | Explore scalable deep learning architectures and integration into continuous monitoring in Zero-Trust environments |
| Le, Zincir-Heywood & Heywood (2020) | ML-based user-centered detection with multi-level data granularity | Proprietary datasets with realistic insider scenarios | Detected 85% malicious insiders at 0.78% FPR; as early as 14 minutes after malicious action | Imbalanced data, limited ground truth, handling behavior drifts | Develop adaptive ML models for concept drift and extend to streaming real-time systems |
| Singh, Mehtre & Sangeetha (2019) | Ensemble Hybrid ML: Multi-State LSTM + CNN for spatio-temporal anomaly detection | Public insider threat datasets | AUC = 0.9042 (train), 0.9047 (test) | Computational complexity; still limited by handcrafted features | Extend ensemble learning with self-supervised learning for continuous insider threat detection |
| Hu et al. (2019) | Continuous user authentication via mouse biobehavioral features with deep learning | Open-source dataset (10 users) | Authentication every 7s; FAR = 2.94%, FRR = 2.28% | Limited to small dataset; not generalizable | Apply to multi-modal behavioral biometrics (keyboard, keystrokes, network activity) in Zero-Trust settings |
| Hall et al. (2018) | Meta-classifier ensemble (boosted ML models) for detecting data exfiltration | CERT r4.2 | ACC = 96.2%, AUC = 0.988 | Focused on a single insider threat scenario; limited generalizability | Expand to multi-scenario detection and integrate with Zero-Trust security policies |
| Gamachchi & Boztas (2017) | Attributed graph clustering + outlier ranking for detecting suspicious users | Enterprise network logs | AUC = 0.7648 | High-dimensional heterogeneous data, scalability issues | Improve graph-based analytics with deep graph neural networks (GNNs) for insider detection |

## III. METHODOLOGY

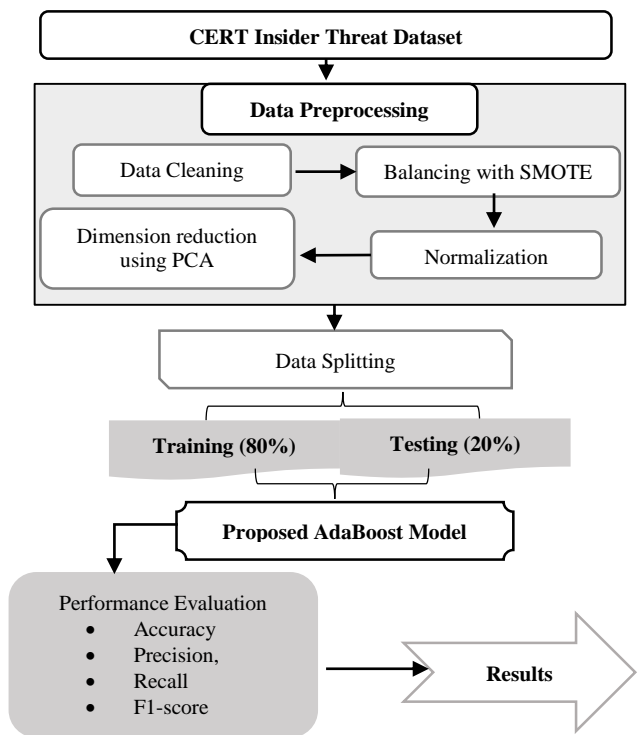

Fig. 1. Proposed Methodology for Insider Threat Detection using AdaBoost Classifier

This study presents a behavioural analytics system that may be used by ZTA to continuously detect insider threats using the CERT Insider Threat Dataset. The methodology starts with an organized pre-processing, where the data cleaning undertaken to remove noise and inconsistencies. To deal with the issue of the imbalance between classes, the SMOTE is utilized, and normalization is utilized to bring the values of features to the same level. PCA is then used to reduce dimensionality, in order to hold vital patterns, whilst reducing redundancy. The data is further sub-divided into training (80%) and testing (20%) data after preprocessing. An AdaBoost classification model is constructed based on the training data and it incorporates a combination of many weak learners to enhance the effectiveness and strength of detection. The trained model is tested against the test data using the measures of performance, which include ACC, PRE, REC and F1. The results demonstrate that AdaBoost is capable of detecting insider threats. As seen in Figure 1, the sequential process.

### A. Data Collection

The CERT Insider Threat database is a voluminous database that is targeted at detection and analysis of insider threat in organizational contexts. It consists of 693,649 records and 830 features and it encompasses a wide range of user activities, behaviors and system interactions that could be an indicator of potential insider malicious activities. The target variable is the insider, which is determined as an indicator of whether a given record was linked with insider threat event. There are numerous different visualizations that give insights into the dataset and are presented below:

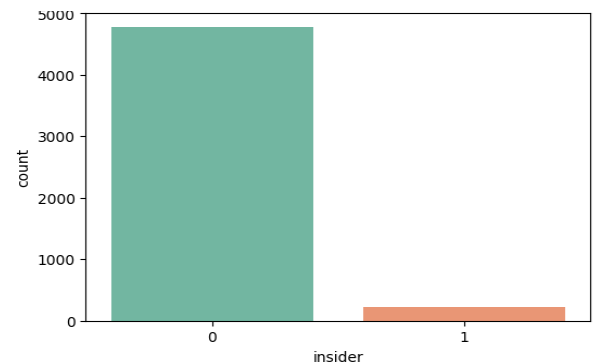

Fig. 2. Insider vs Non-Insider Threat Class Distribution

Figure 2 shows the number of observations of two different categories, which are denoted as insider on the x-axis. There are a lot more representatives of the category entitled 0 (probably, non-insiders), and the number is about 4,800. In sharp contrast, the section marked with a 1 (representing insiders) has much less amount, about 200-300. Such a strong disparity points to the significant imbalance in the data, with the data on non-insider observations drastically exceeding the data on insider observations. This type of distribution is characteristic of classification problems that are highly imbalanced, and which might need special treatment in a data analysis or machine learning task.

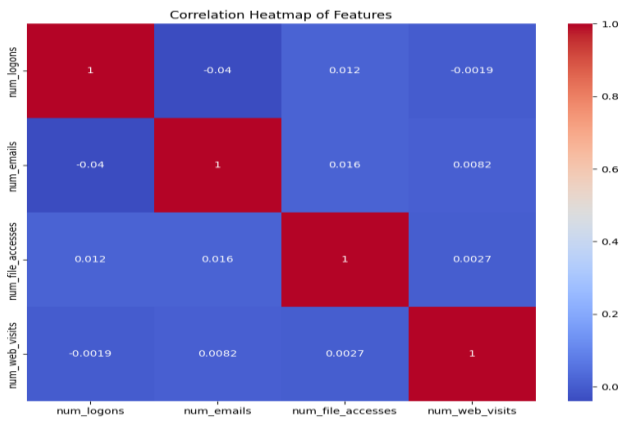

Fig. 3. Correlation Heatmap of the Dataset

The Figure 3, Correlation Heatmap of Features, is a visualization of the four features: num_logons, num_emails, num_file_accesses, and num_web_visits, based upon the CERT insider threat data. The diagonal depicts the perfect correlation of 1 since there is no feature that is not correlated with itself, and all the features are correlated. All off-diagonal values are near to zero showing a very weak or negligible linear relationship existing between any two different features. An example of this is that num_logons and num_emails have a correlation of only -0.04. This implies that the features are mostly independent but this is a highly preferred attribute of machine learning models.

### B. Data Preprocessing

Data Pre-processing is a critical stage of working with raw data to prepare it and be processed by machine learning. It takes care of cleaning, transforming and organizing the data in a manner that is computationally efficient. Proper pre-processing is necessary to ensure the dataset is accurate, consistent, and well organized in order to enhance the reliability and the usability of the dataset in the subsequent model. The steps below provide an outline of pre-processing:

#### 1) Data Cleaning

Pre-processing required is data cleaning, which ensures the quality, reliability, and usefulness of the dataset before running ML algorithms [18]. The proposed study conducted data cleaning checks in the CERT insider threat dataset. Even though no missing or unusual values were found in the dataset, the overall procedure was as follows:

- The data was tested on any of the 830 features containing the null or missing values and none of the cases having the missing values were observed.
- Outliers were verified in all the numerical variables and categorical variables and there were no serious outliers.
- The dataset was checked on the existence of duplicate records and no such records were found.
- The consistency of data in any feature (timestamps, user IDs, and activity logs, etc.) was checked and the data in the dataset had the same formats and values.

### C. Balancing using SMOTE

ML researchers use the SMOTE to solve the issue of inequality in social classes. With the creation of synthetic instances for the minority class, SMOTE strives towards its primary objective of a more equitable distribution of classes. Here is how the sampling procedure works At the outset, an instance of a minority class is selected. Then, after selecting an instance, the algorithm finds its KNN, where k=3 in this study [19]. One neighbour is chosen at random from the list of those who have been identified. The last thing to do is combine the characteristics of the first instance with those of the neighbour that was chosen at random, in order to create a synthetic instance. This practice effectively introduces synthetic. Insider threads 0 and 1 have their number class designations modified to 6,92,342 after SMOTE.

### D. Normalization

Classification techniques that use distance measurements or neural networks, including nearest-neighbour and clustering algorithms, benefit greatly from normalisation because it speeds up the model training stage [20]. Some of the most popular methods for normalisation are min-max, z-score, and decimal scaling. This study makes use of Min-max normalization, which has the following general Equation (1):

$$x_{new} = \frac{x - \min(x)}{\max(x) - \min(x)} \quad (1)$$

### E. Dimension Reduction Using PCA

Dimension reduction algorithms are employed to tackle problems caused by high-dimensional data. This type of data not only makes computation more complex and increases the real-time requirement, but it also hinders fault diagnosis performance because many dimensions are irrelevant and can hide clusters in noisy data [21]. PCA can be used to reduce the dimensions of a data set by converting the correlated features to smaller sized sets of uncorrelated principal components. This accelerates the process of training and increases interpretability of the model with minimal variance loss of the data. Covariance matrix Σ should be calculated first in order to obtain the matrix for projection, as per the Equation (2), for dimension reduction.

$$\Sigma = \frac{1}{m} \sum_{i=1}^{m} (X^{(i)})(X^{(i)})^T \quad (2)$$

### F. Data Splitting

The dataset was split into a training set and a testing set so that we could test our models thoroughly. The data was divided as follows: 80% for training and 20% for testing. The distribution of classes within subsets was preserved using stratified sampling. Building models and doing cross-validation were both done using the training set. Using the testing set, we were able to see how well the final model performed on novel data.

### G. Proposed AdaBoost Model

Adaptive boosting, or AdaBoost [22], was first proposed in 1996 by Freund and Schapiro. Transforming the weak classifier into the strong classifier is the job of AdaBoost. This is a binary classification problem that it is solving. Primarily, Ad boost is used for boosting algorithm construction. A small decision tree was employed by Ad boost. The tree's performance on each training instance is utilised immediately after generation. It is difficult to forecast training data that benefit from additional weights, therefore must first decide how much attention to give to the next level of the tree. To reliably anticipate out-of-the-ordinary observations, Ad boost trains the data sample iteratively while adjusting the classifier weights [23]. To make the classifier perfect, and need train it interactively on many weighted training samples. Each time around, it strives to give the greatest possible match for these samples by reducing training error. Each weak learner in AdaBoost is allocated a contribution (weight) based on the its performance on the weighted training data, and is computed by Equation (3):

$$\alpha_t = \frac{1}{2} \ln\left(\frac{1-\epsilon_t}{\epsilon_t}\right) \quad (3)$$

such that the learners with lower error rates contribute to the final decision at a higher rate. The error of any weak learner is weighted as shown in the Equation (4):

$$\epsilon_t = \sum_{i=1}^{N} w_i^{(t)} \mathbb{1}(h_t(x_i) \neq y_i) \quad (4)$$

In which misclassified samples add more to the error, hence attracting the next iteration to work on harder examples. Lastly, the strong classifier is assembled by aggregating all the weak learners as expressed in Equation (5):

$$F(x) = sign\left(\sum_{t=1}^{T} \alpha_t h_t(x)\right) \quad (5)$$

The boosted model, which is the model attained by using the power of other weak classifiers.

### H. Evaluation Metrics

Performance, efficacy, and dependability of machine learning models are assessed using assessment measures. To ensure the model is robust and generalizable, as well as to assess its pattern-recognizing and prediction-making abilities, they are invaluable. The metrics used in this research gave a detailed account of the model performance.

#### 1) Confusion Matrix

One way to test a classifier's accuracy in distinguishing between different classes is with a confusion matrix [24]. We can see when the classifier got the data classification right by looking at the TP and TN numbers, and when it got it wrong, we can see the FP and FN values.

- **TP (True Positive):** Data with a positive predictive value and a positive actual value in large quantities.
- **FP (False Positive):** Amount of information that can be used for positively and negatively predictive purposes.
- **FN (False Negative):** Data that is positive and has an actual value, and data that is negative and contains forecasts.
- **TN (True Negative):** Data in enormous volumes with negative expected value and negative actual value.

The following measures are utilised to assess a model's performance:

**Accuracy:** A percentage is used to represent ACC. How well the instances are predicted is what it means [25]. The formula for it is given in Equation (6):

$$Accuracy = \frac{TP+TN}{TP+FP+TN+FN} \qquad (6)$$

**Precision:** One measure of ACC is the proportion of right-answers to total decisions. The result of dividing total TP by the product of total FP and TP represents this. Given by Equation (7)

$$Precision = \frac{TP}{TP+FP} \qquad (7)$$

**Recall:** The percentage of positive TPR divided by the total number of positive evaluations is the REC. It is computed using Equation (8):

$$Recall = \frac{TP}{TP+FN} \qquad (8)$$

**F1-score:** The amount of ACC can be tested using the F-measure. Equation. (9) expresses this as the equilibrium between two competing concepts: sensitivity and PRE:

$$F1\ Score = \frac{2\ (Precision \times Recall)}{Precision \times Recall} \qquad (9)$$

This comprehensive evaluation process enables a detailed assessment of multiple aspects of model performance, supporting the identification and selection of the most suitable and effective models.

### IV. RESULT ANALYSIS AND DISCUSSION

The proposed approach emphasizes continuous behavioural analytics for insider threat detection within ZTA. To evaluate the effectiveness of the AdaBoost model, experiments were conducted using a comprehensive insider threat dataset simulating user activity in a zero-trust environment. The experimental setup utilized a workstation equipped with an Intel Xeon W-2295 CPU @ 3.0 GHz and 32 GB of RAM, with models implemented in Python using scikit-learn and relevant machine learning libraries. The results of the AdaBoost model are shown in Table II to detect continuous insider threats. The model has a high ACC of 98.0% which reflects excellent overall predictive ability in detecting possible insider threats. At 98.3-PRE and 98.0-REC, the AdaBoost model has a balanced performance of accurately classifying malicious behaviour and limiting the number of FP and FN. The fact that the F1 is 98.0% is also another indicator that it is capable of addressing the complexities of sequential and behavioural security data. Such findings highlight the AdaBoost model as appropriate to continuous monitoring and threat mitigation in zero-trust security models.

TABLE II.   PERFORMANCE OF THE ADABOOST MODEL FOR INSIDER THREAT DETECTION IN ZERO TRUST ARCHITECTURE

| Matrix | AdaBoost |
|---|---|
| Accuracy | 98.0 |
| Precision | 98.3 |
| Recall | 98.0 |
| F1 Score | 98.0 |

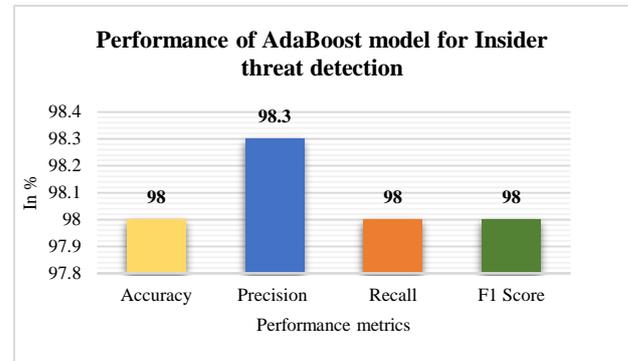

Fig. 4.  Performance of Adaboost Model for Insider Threat Detection

Table II and Figure 4 illustrate the AdaBoost model's classification task performance. The model achieves a high accuracy of 98.0%, indicating it correctly predicts the majority of instances. PRE is the highest at 98.3%, reflecting its strong ability to identify positive cases while minimizing false positives. REC and F1 are both 98.0%, demonstrating robust detection of actual positive instances and a balanced trade-off between PRE and REC. The figure illustrates these metrics on a percentage scale from 97.8% to 98.4%, highlighting AdaBoost's consistently high and reliable performance for insider threat detection.

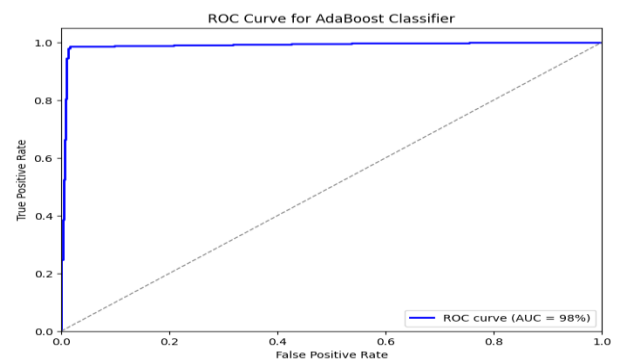

Fig. 5.  ROC Curve of the AdaBoost Model

The ROC curve for the AdaBoost Classifier is displayed in Figure 5. In this graph, shows the relationship between two variables: the sensitivity (or TPR) and the specificity (or FPR ) divided by 1. The blue performance curve of the model is very near to the top left corner of the plot, which indicates excellent outcomes.  An extremely high Area Under the Curve (AUC) of 98% (or 0.98), as is expressly stated, is achieved. An AUC of 0.50 indicates a random classifier, as seen by the dashed diagonal line.  It is clear that the AdaBoost model does a great job of differentiating between the two groups because of the steep ascent of the curve and its closeness to the ideal point (0,1).

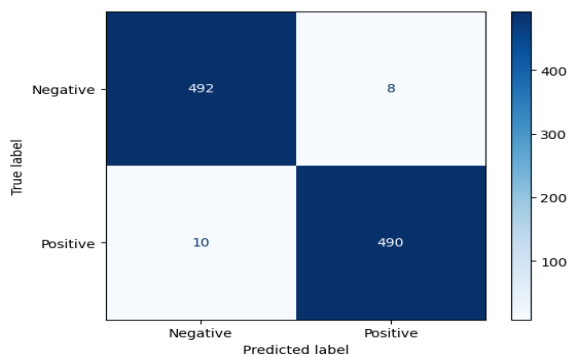

Fig. 6. Confusion Matrix of the AdaBoost Model

Figure 6, a blue heatmap, displays the confusion matrix for a binary classification model. The matrix's vertical axis displays the True label, while the horizontal axis displays the anticipated label. Both labels are classified as Positive or Negative. There is 492 TN (instances of Negative accurately identified) and 8 False Positives (instances of Negative wrongly projected as Positive) in the top-left cell. On the one hand, shows 10 erroneous negative predictions (positive cases) in the lower-left corner, and on the other, 490 correctly classified positive examples (true positives) in the lower-right corner.

A. *Comparison and Discussion*

Table III displays the results of an accuracy-based evaluation of various ML models for continuous insider threat detection in zero-trust infrastructures. When it comes to differentiating between benign and malevolent insider operations, the SVM managed an accuracy of 90.10 percent, suggesting moderate predictive power. The ANN slightly improved performance with 94.74% accuracy, reflecting stronger pattern recognition. Similarly, the Bayesian Network (BaysNet) attained 94.87% accuracy, offering a comparable balance of detection ability. In contrast, the AdaBoost model significantly outperformed all others, reaching 98.0% accuracy, demonstrating its superiority for behavioral analytics and continuous threat detection within zero-trust security environments.

TABLE III.  COMPARATIVE ACCURACY OF MACHINE LEARNING MODELS FOR INSIDER THREAT DETECTION

| Models | Accuracy |
|---|---|
| BaysNet[26] | 94.87 |
| SVM[27] | 90.10 |
| ANN[28] | 94.74 |
| AdaBoost | 98.0 |

The findings show that AdaBoost offers a better method of detection of insider threats than the conventional models like SVM, ANN, and Bayesian Network. Its ensemble structure enables it to record small deviations of behaviour, which would be missed by other models. Comparative analysis reveals that traditional classifiers though good in recognizing overall trends, do not respond to subtle anomalies found in insider activities. The results reveal that behavioural analytics should be combined with adaptive approaches to machine learning in zero-trust systems. Being able to track user activity allows organizations to prevent any insider threat in advance, as well as enhance the overall system security and resilience.

V. CONCLUSION AND FUTURE SCOPE

Behavioural analytics integrated with machine learning proved highly effective for continuous insider threat detection in zero-trust architectures. Pre-processing methods such as data cleaning, SMOTE, normalization, and PCA ensured high-quality feature representation. Comparative experiments revealed that AdaBoost significantly outperformed traditional models. While SVM (90.1% ACC), ANN (94.7%), and Bayesian Network (94.9%) achieved moderate success, AdaBoost surpassed them with 98.0% ACC, 98.3% PRE, 98.0% REC, and an F1 of 98.0%. The ROC analysis confirmed an AUC of 0.98, demonstrating superior discriminative power and balanced performance. These results highlight AdaBoost's strength in detecting subtle behavioural anomalies while minimizing false positives and negatives, making it well-suited for dynamic zero-trust environments. Future research can focus on real-time deployment to validate scalability under enterprise-scale workloads. Expanding beyond the CERT dataset, multi-modal data such as keystroke dynamics, biometric signals, and network patterns can enrich behavioural profiling. Advanced learning techniques, including graph NN, RL, and self-supervised approaches, may enhance adaptability to evolving threats. Incorporating explainable AI is another critical step, providing transparency in model outputs to improve analyst trust and actionable insights. Together, these directions can advance insider threat detection into a proactive, scalable, and interpretable solution for zero-trust security frameworks.

REFERENCES

[1] S. F. Ahmed and N. A. Hikal, "A Review of Cyber-security Measuring and Assessment Methods for Modern Enterprises," *JOIV Int. J. Informatics Vis.*, vol. 3, no. 3, pp. 157–176, Aug. 2019, doi: 10.30630/joiv.3.3.241.

[2] S. Neeli, "The Significance of NoSQL Databases : Strategic Business Approaches and Management Techniques," *J. Adv. Dev. Res.*, vol. 10, no. 1, p. 11, 2019.

[3] J. Modini, M. Vanzomeren, S. Fowler, K. Joiner, and T. Lynar, "Rising to the Challenge of Insider Threats for Middle Powers," in *International Conference on Cyber Warfare and Security*, 2020. doi: 10.34190/ICCWS.20.131.

[4] P. Pathak, A. Shrivastava, and S. Gupta, "A Survey on Various Security Issues in Delay Tolerant Networks," *J. Adv. Shell Program.*, vol. 2, no. 2, pp. 12–18, 2015.

[5] V. M. L. G. Nerella, "Automated cross-platform database migration and high availability implementation," *Turkish J. Comput. Math. Educ.*, vol. 9, no. 2, pp. 823–835, 2018.

[6] A. Balasubramanian, "Proactive Machine Learning Approach to Combat Money Laundering in Financial Sectors," *Int. J. Innov. Res. Eng. Multidiscip. Phys. Sci.*, vol. 7, no. 2, pp. 1–15, 2019, doi: 10.5281/zenodo.14508474.

[7] M. K. Omopariola, "Zero-Trust Architecture Deployment in Emerging Economies: A Case Study from Nigeria," *Int. J. Comput. Appl. Technol. Res.*, vol. 5, no. 12, 2016.

[8] A. Balasubramanian, "Ai-Enabled Demand Response: A Framework For Smarter Energy Management," *Int. J. Core Eng. Manag.*, vol. 5, no. 6, pp. 96–110, 2018.

[9] D. S. Berman, A. L. Buczak, J. S. Chavis, and C. L. Corbett, "A Survey of Deep Learning Methods for Cyber Security," *Information*, vol. 10, no. 4, 2019, doi: 10.3390/info10040122.

[10] S. S. S. Neeli, "Serverless Databases: A Cost-Effective and Scalable Solution," *Int. J. Innov. Res. Eng. Multidiscip. Phys. Sci*, vol. 7, no. 6, p. 7, 2019.

[11] M. Zamani and M. Movahedi, "Machine learning techniques for intrusion detection," *arXiv Prepr. arXiv1312.2177*, 2013.

[12] T. Tamanna, "Detection of insider threats based on deep learning using LSTM--CNN model," Dublin, National College of Ireland, 2020.

[13] D. C. Le, N. Zincir-Heywood, and M. I. Heywood, "Analyzing Data Granularity Levels for Insider Threat Detection Using Machine Learning," *IEEE Trans. Netw. Serv. Manag.*, vol. 17, no. 1, pp. 30–44, 2020, doi: 10.1109/TNSM.2020.2967721.

[14] M. Singh, B. M. Mehtre, and S. Sangeetha, "User Behavior Profiling using Ensemble Approach for Insider Threat Detection," in *ISBA 2019 - 5th IEEE International Conference on Identity, Security and Behavior Analysis*, 2019. doi: 10.1109/ISBA.2019.8778466.

[15] T. Hu, W. Niu, X. Zhang, X. Liu, J. Lu, and Y. Liu, "An Insider Threat Detection Approach Based on Mouse Dynamics and Deep Learning," *Secur. Commun. Networks*, vol. 2019, 2019, doi: 10.1155/2019/3898951.

[16] A. J. Hall, N. Pitropakis, W. J. Buchanan, and N. Moradpoor, "Predicting Malicious Insider Threat Scenarios Using Organizational Data and a Heterogeneous Stack-Classifier," in *2018 IEEE International Conference on Big Data (Big Data)*, 2018, pp. 5034–5039. doi: 10.1109/BigData.2018.8621922.

[17] A. Gamachchi and S. Boztas, "Insider Threat Detection Through Attributed Graph Clustering," in *2017 IEEE Trustcom/BigDataSE/ICESS*, 2017, pp. 112–119. doi: 10.1109/Trustcom/BigDataSE/ICESS.2017.227.

[18] S. S. S. Sindhu, S. Geetha, and A. Kannan, "Decision tree based light weight intrusion detection using a wrapper approach," *Expert Syst. Appl.*, vol. 39, no. 1, pp. 129–141, 2012.

[19] D. Gonzalez-Cuautle *et al.*, "Synthetic Minority Oversampling Technique for Optimizing Classification Tasks in Botnet and Intrusion-Detection-System Datasets," *Appl. Sci.*, vol. 10, no. 3, p. 794, Jan. 2020, doi: 10.3390/app10030794.

[20] I. A. Khan, D. Pi, Z. U. Khan, Y. Hussain, and A. Nawaz, "HML-IDS: A Hybrid-Multilevel Anomaly Prediction Approach for Intrusion Detection in SCADA Systems," *IEEE Access*, vol. 7, pp. 89507–89521, 2019, doi: 10.1109/ACCESS.2019.2925838.

[21] U. A. Korat and A. Alimohammad, "A Reconfigurable Hardware Architecture for Principal Component Analysis," *Circuits, Syst. Signal Process.*, vol. 38, no. 5, pp. 2097–2113, May 2019, doi: 10.1007/s00034-018-0953-y.

[22] A. B. Abhale and S. S. Manivannan, "Supervised Machine Learning Classification Algorithmic Approach for Finding Anomaly Type of Intrusion Detection in Wireless Sensor Network," *Opt. Mem. Neural Networks*, vol. 29, no. 3, pp. 244–256, Jul. 2020, doi: 10.3103/S1060992X20030029.

[23] B. Mueller, "Enhancing Hidden Threat Detection in Cybersecurity Using Machine Learning Technology and Information Security Event Management (SIEM)," 2020.

[24] A. Yulianto, P. Sukarno, and N. A. Suwastika, "Improving AdaBoost-based Intrusion Detection System (IDS) Performance on CIC IDS 2017 Dataset," *J. Phys. Conf. Ser.*, vol. 1192, Mar. 2019, doi: 10.1088/1742-6596/1192/1/012018.

[25] S. Zhao, W. Li, T. Zia, and A. Y. Zomaya, "A Dimension Reduction Model and Classifier for Anomaly-Based Intrusion Detection in Internet of Things," in *2017 IEEE 15th Intl Conf on Dependable, Autonomic and Secure Computing, 15th Intl Conf on Pervasive Intelligence and Computing, 3rd Intl Conf on Big Data Intelligence and Computing and Cyber Science and Technology Congress(DASC/PiCom/DataCom/CyberSciTech)*, IEEE, Nov. 2017, pp. 836–843. doi: 10.1109/DASC-PICom-DataCom-CyberSciTec.2017.141.

[26] D. C. Le, S. Khanchi, A. N. Zincir-Heywood, and M. I. Heywood, "Benchmarking evolutionary computation approaches to insider threat detection," in *Proceedings of the Genetic and Evolutionary Computation Conference*, Jul. 2018, pp. 1286–1293. doi: 10.1145/3205455.3205612.

[27] O. Almomani, "A Feature Selection Model for Network Intrusion Detection System Based on PSO, GWO, FFA and GA Algorithms," *Symmetry (Basel).*, vol. 12, no. 6, 2020, doi: 10.3390/sym12061046.

[28] M. Al-Zewairi, S. Almajali, and M. Ayyash, "Unknown Security Attack Detection Using Shallow and Deep ANN Classifiers," *Electronics*, vol. 9, no. 12, 2020, doi: 10.3390/electronics9122006.